\documentstyle[11pt,newpasp,twoside,astrobib,psfig]{article} 
\def\edcomment#1{\iffalse\marginpar{\raggedright\sl#1\/}\else\relax\fi} 
\reversemarginpar 

\relax
\begin{document} 
\title{Blazars in Low-Luminosity and Radio-Weak AGN?} 
 \author{Heino Falcke, Sera Markoff, Peter L. Biermann} 
\affil{Max-Planck-Institut f\"ur Radioastronomie, Auf dem H\"ugel,
D-53121 Bonn, Germany} 
\begin{abstract} 
  Typical blazars seem to be associated with FR I and FR II radio galaxies and
radio-loud quasars. However, what happens at lower powers? Do blazars exist in
low-luminosity AGN or do they exist in radio-quiet AGN? A recent detection of
superluminal motion in a supposedly radio-quiet Seyfert raises the question
whether beaming can play an important role in some of these objects as well.
Moreover, VLBI observations of nearby low-luminosity AGN reveal compact
flat-spectrum radio cores very similar to those in bright radio-loud blazars.
Furthermore, with the detection of X-ray emission from the least luminous AGN
we can study, Sgr A* in the Galactic Center, this source now seems to be
dominated entirely by non-thermal emission -- like in BL Lacs. The same may be
true for some X-ray binaries in the Low/Hard-state. Inclusion of low-power
radio jets into the overall picture of AGN provides some clues for what type of
accretion is important, what the power, radiative efficiency and matter-content
of jets is, and what mechanism could be responsible for making jets radio-loud.
We specifically discuss whether proton-proton collisions in a hot accretion
flow could provide the switch for the radio-dichotomy.
\end{abstract} 
\section{Introduction} 
Over the last decades our basic picture of what blazars are has
solidified. It is not even discussed anymore that the cause of the
blazar and BL Lac phenomenon are relativistic jets and their
non-thermal emission. In many cases this jet emission dominates the
spectrum of these sources over the entire electromagnetic spectrum --
from radio through TeV. The main discussions today revolve mainly
around the internal structure and parameters of the jets, and the
particle processes responsible for the emission.

Within the blazar community we have become so accustomed to jets that
we take them for granted, however, outside the community jets are
still considered something exotic: one tries to get away with thermal,
spherical or disk-only models as long as possible. One example
certainly is the Gamma-Ray Burst (GRB) field. This are the most
violent, non-thermal events we know in the universe. After all the
experience with the history of AGN research, it should have been
natural to start with the assumption that these energetic sources have
a setting similar to the other violent sources we know in the universe
-- blazar jets. Still, only after many years and under the
overwhelming weight of extremely luminous GRBs like GRB971214, the
fireball model collapsed into various jet models (one of which was
developed by us and which we cannot resist to reference,
i.e.~\citeNP{PuglieseFalckeBiermann1999}).

Another example are X-ray binaries, which are considered micro-quasars
\cite{MirabelRodriguez1999} and where many jets have been observed
\cite{FenderHendry2000,FenderKuulkers2000}.  However, so far we have
not identified anything similar to a blazar or BL Lac in this field,
and most emission models seem to ignore completely the non-thermal jet
component. Is this justified in all cases?

Finally, we need to mention the AGN-field itself. Clearly, blazars are
just the tip of the iceberg since we usually are only looking at the
most luminous AGN.  Only some 10\% of these can be considered
radio-loud, and of these only a small fraction will point towards us to
reveal the amplified non-thermal spectrum through boosting. Of course
there are many more AGN around: radio-quiet quasars and low-luminosity
AGN that we never really consider to be blazars or blazar-like, yet
even here jets are important ingredients.

Thus, the question we want to address is: can we identify blazars or
blazar-like sources in other classes of astrophysical sources? How do
we find them and what are we looking for? In the end this boils down
to the question what the definition of a blazar actually is. Does
``blazar'' in this context imply an observational classification or a
description of a physical state? For example, in luminous AGN
relativistic boosting is very important since it makes it possible for
the non-thermal emission to dominate over any thermal emission. But
what happens if for some reason the thermal emission is strongly
suppressed anyway, e.g. by a hot, radiatively inefficient accretion
flow? In this case we will still see the non-thermal jet emission
dominating even from large angles of the line-of-sight to the jet axis
and we may still have substantial variability -- do we call this a
blazar even without relativistic boosting being important? And if not,
how are we going to tell such a source apart from a boosted jet in
surveys? Determining the Doppler factor of jets is still very
difficult.

Rather than answering all these question here, we will rather discuss
a few recent observational results on non-traditional blazars sources,
that nevertheless show how difficult such a classification can
become. Since some of the distinctions among AGN are related to the
question of radio-loudness we will also discuss this issue here, since it
was hotly debated at this workshop.

\section{Radio Dichotomy}
The first quasars were actually discovered through their radio identification
\cite{HazardMackeyShimmins1963} and belonged to the blazar class. It 
was soon found that many other quasars (QSOs) exist that do not show
such strong radio emission. Indeed, support for a bimodal distribution
of radio luminosities (relative to optical luminosity) came from radio
observations of optically
\cite{StrittmatterHillPauliny-Toth1980,KellermannSramekSchmidt1989,FalckeSherwoodPatnaik1996} 
and X-ray selected quasar samples
\cite{dellaCecaLamoraniMaccacaro1994}. In some cases with lower
statistics the evidence for an actual bimodality remained ambiguous
\cite{HooperImpeyFoltz1996} and White (this conference, see also
\citeNP{HelfandBeckerGregg1999}) claimed that in the (radio-selected) 
FIRST quasar survey a bimodality was not confirmed. Similarly, at this
conference Meg Urry raised the question whether in fact there is a
continuous distribution of radio-to-optical luminosities rather than a
dichotomy. Are we fooled similar to the apparently artificial
distinction between X-ray and radio-selected BL Lacs that is now
becoming replaced by a continuum of sources just peaking at different
frequencies?

The answer to this challenge is probably that both camps are not
entirely wrong, depending on what the actual question is. The problem
is, that any dichotomy in the physics of AGN and of jet formation can
be easily washed out by various effects. A survey that contains a
mixed bag of AGN -- old and young, luminous and faint, obscured and
face on -- may in fact not show any bimodality while the underlying
physics still is. Why is this?

The dichotomy is seen so far mainly in optically selected samples of
quasars with rather homogeneous properties: strong and luminous UV
bump and broad emission lines. Assuming this is indeed emission from
radiatively efficient accretion disks, the optical luminosity should
be a good measure of the accretion power onto the central black
hole. Similarly, if the radio properties of quasars are homogeneous
then the radio-luminosity can be a useful estimate for the jet
power. A constant radio-to-optical flux density ratio (the
$R$-parameter) then reflects nothing else but a constant fraction of
the accretion power being channeled into a jet. This is one of the
motivations behind the jet-disk symbiosis idea
\cite{FalckeBiermann1995} and explains the radio optical correlations
of quasars
\cite{BaumHeckman1989,MillerRawlingsSaunders1993,FalckeMalkanBiermann1995}.
A bimodality in the $R$-parameter than implies a dichotomy in the jet
formation -- whatever that mechanism is.

On the other hand the $R$-parameter is a rather shaky measure for such
a physical effect. First of all, the radiative efficiency of jets need
not be constant and in fact depends on many external factors. Doppler
boosting of the core certainly affects it, and for hotspots the
pressure of the surrounding medium and their location of the jet
terminus (inside or outside the galaxy) can change the radio output
quite strongly for a fixed jet power. The latter is seen for example
in the declining radio power for jets as they grow from GPS
(Gigahertz-Peaked-Spectrum) to CSS (Compact-Steep-Spectrum) sources
and finally evolve into large radio galaxies
(e.g.,~\citeNP{O'Dea1998}).

\begin{figure}
\centerline{\psfig{figure=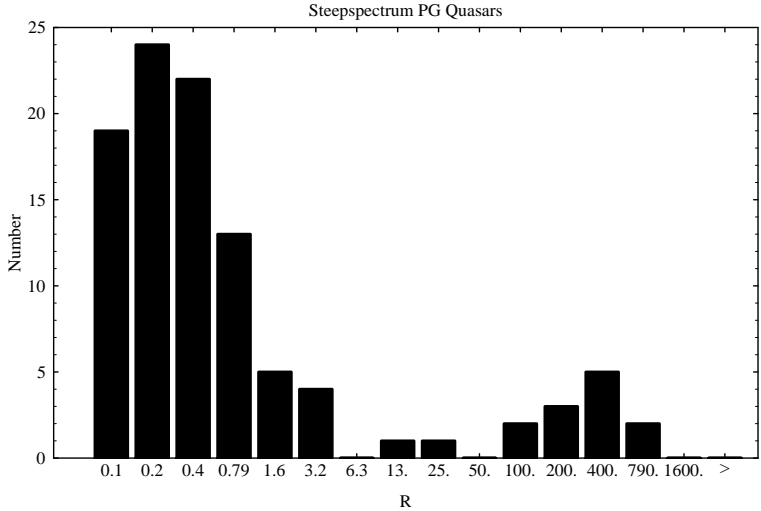,width=0.75\textwidth,bbllx=2.6cm,bblly=8.6cm,bbury=19.7cm,bburx=19cm}}
\caption[]{\label{dichotomy} Distribution of the radio-to-optical flux 
density ratio $R$ for steep-spectrum PG-quasars (from Falcke,
Sherwood, Patnaik 1996)}
\end{figure}

On the other hand, the optical luminosity can easily be obscured or
could be entirely absent so that a reliable estimate of the accretion
power is not possible. The optical selection of the PG quasar sample,
on which many of the papers mentioned above are based on, seemed to
have avoided many of these problems. In this context ``bias'' can be
good since it provides one with a rather clean sample that gives us
some insight into the physics of jet formation. This is illustrated in
Figure~\ref{dichotomy} where we show the distribution of
$R$-parameters in the PG quasar sample, however, with all known
flat-spectrum ($\alpha>-0.5$ at 5 GHz) sources taken out -- this
discriminates strongly against those quasars affected by boosting and
also against GPS sources. The resulting distribution is rather clean
and bimodal (but is also present if one does not take out the
flat-spectrum quasars).

The answer to the question whether jet formation in luminous quasars
is bimodal, is therefore most likely ``yes''. On the other hand, the
question whether the $R$-parameter distribution for `all the AGN on the
sky' is bimodal, is most likely ``no''.

\nocite{FalckeSherwoodPatnaik1996}

\section{Radio-Quiet Quasars \& Seyferts}
One interesting finding from studying the $R$-parameter distribution
in PG quasars was that the region intermediate between the radio-loud
and radio-quiet distribution was not empty but populated with many
flat-spectrum sources
\cite{MillerRawlingsSaunders1993,FalckeMalkanBiermann1995,FalckeSherwoodPatnaik1996},
called radio-intermediate quasars (RIQs). Some of these sources had
radio properties similar to those of blazars: core-dominated,
flat-spectrum, and variable. The suggestion was that they constitute a
population of relativistically boosted radio-quiet quasars,
i.e.~something one might call radio-weak blazars.

\begin{figure}
\centerline{\psfig{figure=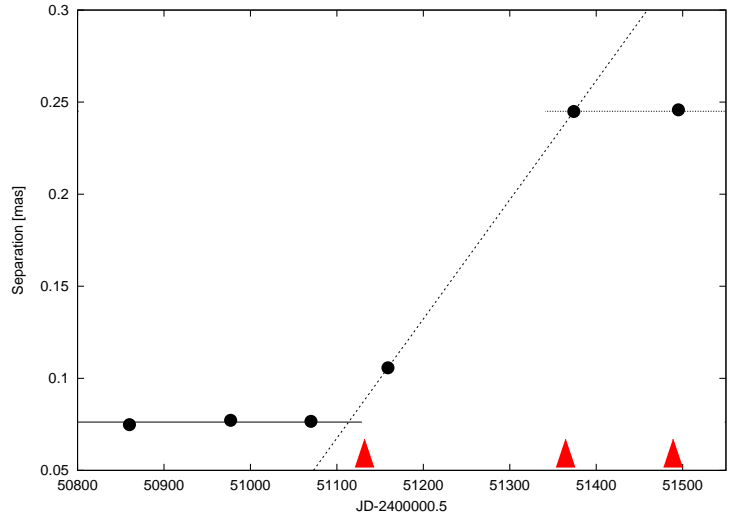,width=0.75\textwidth,angle=180}}
\caption[]{\label{iiizw2} Component separation as a function of time
for the Seyfert galaxy III~Zw~2 as measured by the VLBA at 43
GHz. There is a short period of superluminal expansion indicating the
presence of a relativistic jet in this Seyfert galaxy.  (From
Brunthaler, Falcke, Bower et al.~2000)}
\end{figure}

To make this case watertight one would have liked to see superluminal
motion in these sources. Early VLBI observations did not reveal
anything but a compact core. However, one of the galaxies, III~Zw~2,
then became target of a monitoring campaign during a major outburst
starting in 1997.  Barely resolved in the first three epochs it
suddenly started to expand for a brief period of about a few months
\cite{BrunthalerFalckeBower2000} and then held steady while 
continuing to decrease in flux (Fig.~\ref{iiizw2}). The spectral
evolution monitored with the VLA indicated that the expansion happened
on an even shorter time scale and depending on which timescale one
takes the implied expansion speed was between 1.2 and 2.7 $c$,
i.e.~superluminal even in the most conservative case.

\nocite{BrunthalerFalckeBower2000b}

This finding is interesting, since it shows that indeed the RIQs
contain relativistic jets. Apart from the bright radio core, III~Zw~2
has all the properties of a Seyfert galaxy with a spiral host and very
faint, uncollimated, and short radio lobes -- certainly not an FR\,I
or FR\,II radio galaxy. Is this then a radio-weak blazar? It certainly
has all criteria one might require: core-dominance, variability, and
superluminal motion. However, there is one additional factor
complicating things: the evolution of the outburst -- with a stop and
go behavior -- suggests that the jet itself is strongly interacting
with the surrounding ISM. \cite{BrunthalerFalckeBower2000b} suggest
that this could be explained with an `inflating balloon' model and the 
formation of ultra-compact hotspots in this galaxy, similar, yet
smaller in size, to those seen in GPS and CSS sources.

Consequently the large radio flux and compact size is not entirely due
to boosting alone even though a relativistic jet is clearly
present. Here we may be pointed also into another direction, if
III~Zw~2 can be regarded as a luminous and perhaps extreme case of a
Seyfert galaxy -- i.e., a supposedly radio quiet AGN. Like in
radio-loud quasars, the jet might start out relativistically but then
is disrupted already on the sub-parsec scale, possibly interacting
with the torus in some cases, and then propagates outwards
sub-relativistically as seen on parsec-scales with VLBI
\cite{UlvestadWrobelRoy1998,RoyWilsonUlvestad2000} forming the
characteristic uncollimated Seyfert radio lobes
\cite{UlvestadWilson1989}. The denser ISM in spirals compared to
ellipticals may play a role here.

Whether or not the jet-ISM interaction in III~Zw~2 is important, the
relativistic jet suggests that blazar-like radio-quiets should be
detectable, possibly at rather high radio-frequencies. A few attempts
to find them with the VLBA are underway
\cite{BlundellBeasley1998}. An alternative way to look
for radio-weak blazars is to do variability studies. Recently
Barvainis et al. (in prep., see also
\citeNP{FalckeLeharBarvainis2001}) made multi-epoch (11
epochs) observations with the VLA of a sample of 30 radio-quiet,
radio-intermediate, and radio-loud quasars at 8.5 GHz. They found up
to 20\% variability within one year in the cores of radio-quiet and
radio-intermediate quasars exceeding that of cores in (non-blazar)
radio-loud quasars. Even among a sample of ill-selected core-dominated
radio-sources (used as phase-calibrators but dominated by blazars)
only a few sources showed somewhat more variability on the same
timescale. Clearly, the compact radio-emission in these variable
radio-quiet/intermediate quasars is AGN-dominated and most likely from
a jet. It would be worth investigating some of the top variability
performers (after III~Zw~2 which tops that list in this survey) in
greater detail with the VLBA to look for jets and superluminal motion
again.

\section{Low-Luminosity AGN}
Of course, blazars could be dim not only because they are radio-weak,
but also because the jet power and the accretion rate onto the black
hole are low. Blazars and BL Lacs have been associated with both FR\,I
and FR\,II radio galaxies \cite{UrryPadovani1995} and hence span a
large range of luminosities and jet powers already. Does this continue
further down to even less luminous AGN? \citeN{MarchaBrowneImpey1996}
studied a sample of fainter core-dominated AGN and found a number of
blazar-like sources, however, with a large spread in optical and
line-emission properties that makes it once more difficult to decide
what to call a BL Lac. Clearly, once the power is very low, emission
from the galaxy -- hot gas and stars -- will dilute every blazar
spectrum.

This saga continues at even lower
powers. \citeN{NagarFalckeWilson2000}, \citeN{FalckeNagarWilson2000},
and \citeN{FalckeNagarWilson2000b} studied a sample of Low-Luminosity AGN
(LLAGN), selected from the spectroscopic survey of
\citeN{HoFilippenkoSargent1995} with the VLA and the VLBA.  A large
fraction of these AGN showed flat-spectrum compact radio cores in
their nuclei. In the brightest cores the VLBA resolved the radio
emission into jet-like structures. Morphology and spectral index
together therefore suggest that the compact radio emission, like in
radio-loud and radio-quiet quasars, is produced in an outflow rather
than in an advection dominated accretion flow (ADAF,
e.g. \citeNP{YiBoughn1998}). Finally, a comparison of the radio flux
densities at several epochs revealed rather strong intra-year
variability with peak-to-peak variations of up to 200-300\% in some
cases. X-ray emission is also detected in many of these LLAGN
\cite{TerashimaHoPtak2000} as are optical/UV continuum point
sources \cite{MaozFilippenkoHo1995}. Of course, the UV-to-X-ray
emission might come from an accretion disk, as commonly assumed, but
how can we be sure? Could some of it also be non-thermal emission from
a relativistic jet pointing towards us? How could we possibly pick out
a very weak BL Lac from a `normal' source? 

\begin{figure}
\centerline{\psfig{figure=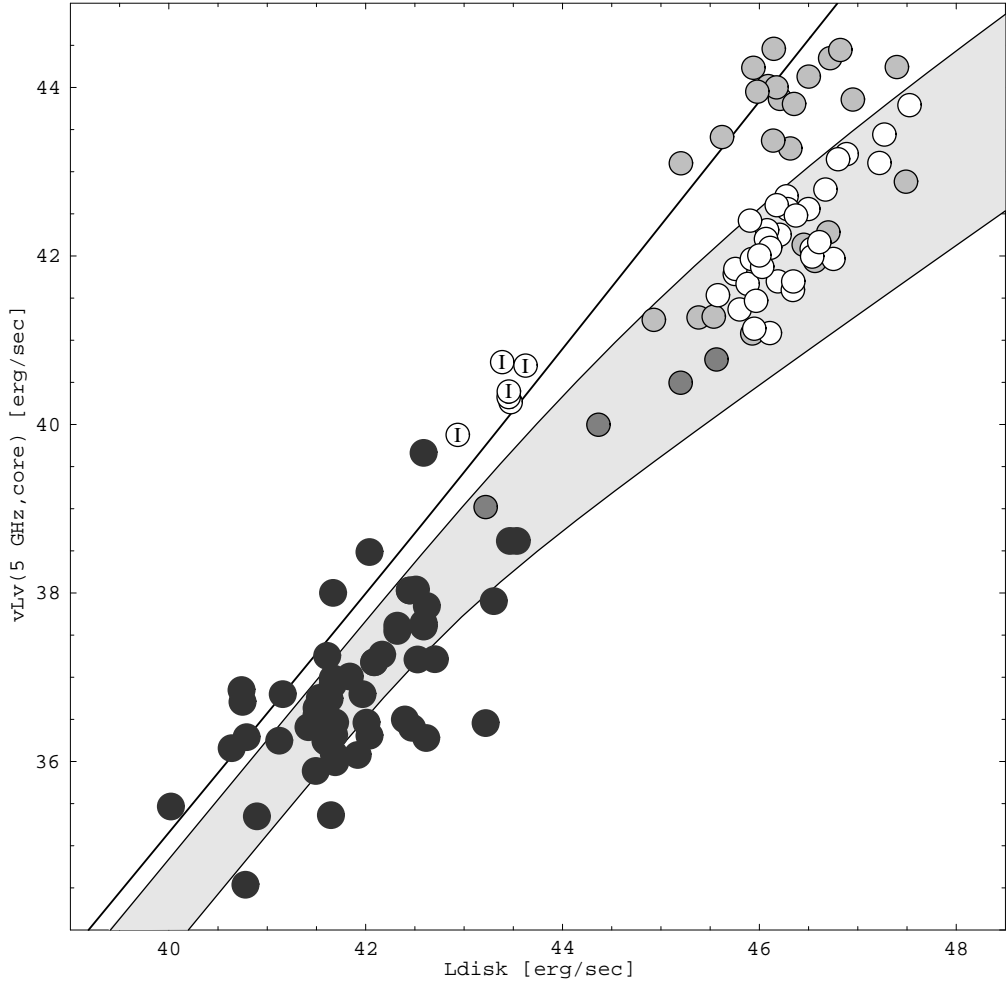,width=\textwidth,bbllx=2.5cm,bblly=5.5cm,bburx=19.1cm,bbury=21.9cm,clip=}}
\caption[]{\label{fr2} 5 GHz radio core vs.~bolometric nuclear luminosity
(derived from narrow H$\alpha$ luminosities) for radio-loud AGN as
given in Falcke, Malkan, Biermann (1995) and Falcke (1996). Solid
black dots are LLAGN from Falcke, Nagar, Wilson (2000). Circles marked
``I'' are FR\,I radio galaxies from Rawlings \& Saunders (1991), open
circles are quasars. Gray shades indicate flat-spectrum,
core-dominated quasars, dark gray shades are radio-intermediate
quasars and Seyferts. The lines are the radio-loud jet-disk symbiosis
model from Falcke \& Biermann (1996), where the thick line represents
sources with inclination angles at the boosting cone, i.e. what is
expected for blazars. Interestingly the LLAGN at this line (black
dots) are all giant ellipticals and seem to connect to FR\,I radio
galaxies. These sources may be underluminous in H$\alpha$ (i.e. $L_{\rm
disk}$) emission and might need to be shifted to the right.}
\end{figure}
\nocite{FalckeMalkanBiermann1995}
\nocite{Falcke1996d}
\nocite{FalckeNagarWilson2000b}
\nocite{RawlingsSaunders1991}
\nocite{FalckeBiermann1996}

This confusion is highlighted in Fig.~\ref{fr2} where we show the
radio-optical correlation for a mixed sample of radio-loud AGN
together with the predictions from the jet-disk symbiosis model from
\citeN{FalckeBiermann1996} outlining regions of constant inclination
angle to guide the eye. Besides an overall scaling of the radio core
power with optical emission it is interesting that the radio cores of
giant elliptical LLAGNs seem to connect to the radio cores in FR\,I
radio galaxies. They are also close to the ``blazar'' line (solid) and
are slightly offset from the rest of the LLAGN population. It is not
clear at present, whether this represents another radio-loudness
dichotomy at lower luminosities or not. There is some evidence that
FR\,Is are simply underluminous in emission lines
\cite{ZirbelBaum1995}, perhaps due to obscuration
(\citeNP{FalckeGopal-KrishnaBiermann1995}, but see
\citeNP{ChiabergeCapettiCelotti2000}) or due to radiatively deficient
disks. This would shift the entire population to the right and thus
below the ``blazar'' line.

Clearly the FR\,Is and their low-luminosity siblings cannot be all
blazars in the sense that they are all pointing towards us. On the
other hand \citeN{ChiabergeCapettiCelotti2000} (also at this workshop)
argue convincingly that if an optical point source is found in an
FR\,I radio galaxy it is probably non-thermal synchrotron
emission. Therefore these sources may be dominated by non-thermal
emission but perhaps are not ``blazing''. In line with the BL
Lac/FR\,I unification scheme one should still also find some sources
that are well above the ``blazar'' line shown in Fig.~\ref{fr2}
because of strong relativistic boosting and suppression of H$\alpha$
emission. So far the sample sizes for LLAGN are probably too small to
find such sources but some of them might have been included in the
\citeN{MarchaBrowneImpey1996} survey.

\section{Sgr A* and the nature of the radio-dichotomy}
In our journey to lower luminosities we now want to take a brief look
at the least luminous AGN we know of, and which may offer some clues
for the physics of jets and blazars.  Sgr A* is now generally believed
to be the central black hole of the Galaxy and its emission mechanism
has been strongly debated (see
\citeNP{MeliaFalcke2001} for a review). However, recent X-ray
observations have provided some interesting new
insight. \citeN{BaganoffBautzBrandt2000} detected a point source at
the position of Sgr A* with Chandra, however, with a rather soft
spectrum -- too soft for thermal bremsstrahlung.  On the other hand
this X-ray spectrum fits nicely with the expectations one has from
synchrotron self-Compton (SSC) emission from the submm-wave emission
region (``submm-bump'') in Sgr A*. The parameters of this bump are
rather well-constrained
\cite{Falcke1996b,BeckertDuschl1997} and hence calculating SSC
emission is straight forward. In fact the entire radio-through-X-ray
spectrum of Sgr A* is now very well fit by pure non-thermal emission
from a radio jet \cite{FalckeMarkoff2000}, the main SSC contribution
coming from the nozzle.

\begin{figure}
\centerline{\psfig{figure=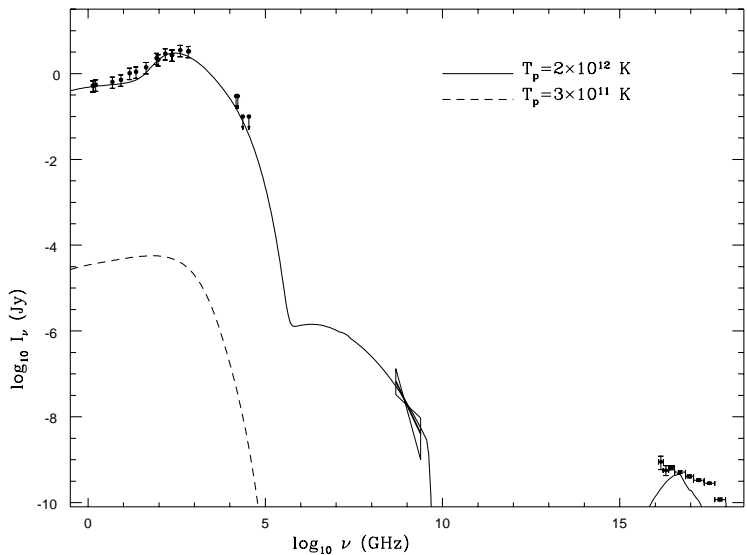,width=0.8\textwidth,angle=-90}}
\caption[]{\label{sgra}Radio through $\gamma$-ray spectrum for Sgr A* resulting from
proton-induced $e^\pm$'s in a hot ($T_p=2\cdot10^{12}$ K) accretion
flow fed into a plasma jet. This includes radio emission from jet and
nozzle, X-ray emission via synchrotron self-Compton in the jet, and
$\gamma$-rays from $\pi^0$-decay in the accretion flow.  The
$\gamma$-ray data are considered upper limits because of the large
beam of the observations.  The fitted parameters for the jet are:
nozzle width $2r_0\sim 5\cdot10^{12}$ cm, height $z_0/r_0=8$,
inclination angle of jet with respect to line-of-sight
$\theta_i=23^\circ$, and $B_0=20$ G (implying an equipartition factor
of $k=0.2$, i.e., magnetically dominated). The fit is very tightly
constrained and the remaining free parameters given here have an
inherent scale, i.e.~gravitational radius or equipartition value.  The
dashed line shows the spectrum for the same parameters but with
$T_p=3\cdot10^{11}$ K.}
\end{figure}

As one can see from the model fit to Sgr A* in Fig.~\ref{sgra} that the
spectrum up to X-rays can be explained by two humps (radio and SSC),
very similar to what is seen in BL Lacs
\cite{FossatiMaraschiCelotti1998}. The absence of any thermal emission
component provides another similarity. A major and interesting
difference, however, is the fact that Sgr A* does not show an
optically thin power-law at high frequencies: the strong IR limits
require an almost exponential cut-off of the synchrotron spectrum. The
absence of such an electron power-law also explains the compactness of
the jet (as seen also in M81; \citeN{BietenholzBartelRupen2000}),
since, in contrast to the flat-spectrum core which is the compact $\tau=1$
surface of the jet, the extended jet emission in AGN is always
optically thin, steep-spectrum emission from a power-law.

In Sgr A* we are very confident that the highest synchrotron
frequencies come from the smallest region, just a few Schwarzschild
radii from the black hole (e.g.,
\citeNP{KrichbaumGrahamWitzel1998}). Hence, the electron population we
see in Sgr A* is probably the freshly injected particle population at
the base of the jet and the inner region of the accretion flow {\it
before} the particles are redistributed by shock
acceleration. Therefore Sgr A* may offer a unique perspective into jet
formation and particle acceleration.  From the frequency of the SSC
peak relative to the synchrotron peak and the shape of the submm-bum
one can then directly derive that the characteristic electron Lorentz
factor has to be around $\gamma_{\rm e}\sim10^2$. This fits nicely
expectations for the minimum Lorentz factor of electrons in radio-loud
quasar jets and hence also in blazars
\cite{CelottiFabian1993,FalckeBiermann1995}.

One can then ask what creates these particles at the base of the jet?
The best explanation for the absence of a thermal bump in the Sgr~A*
spectrum is a hot, optically thin accretion flow
\cite{Rees1982,Melia1992a,NarayanMahadevanGrindlay1998} and some of
the synchrotron emission in the jet could come from the hot electrons
near the inner edge of the accretion disk being advected in a jet. An
alternative proposal \cite{MarkoffFalckeBiermann2000} is that
proton-proton collisions could be responsible. As soon as the protons
in the flow reach a threshold temperature around $10^{12}$ K,
$pp$-collisions become inelastic and inevitably produce pairs,
neutrinos, and $\gamma$-rays in hadronic cascades. The resulting
electron/positron pair-spectrum peaks around 30 MeV, essentially what
is needed for Sgr A*. The pair-production rate is a function of the
accretion rate and the viscosity parameter $\alpha$ of the accretion
flow. For the parameters of Sgr A*, the right number of pairs is
produced for accretion rates of order $\alpha^{-1}\dot
M\sim10^{-4}M_\odot$ yr$^{-1}$, just what is expected from Bondi-Hoyle
accretion of stellar winds.

The spectrum from this process in conjunction with a jet model is what
is actually shown in Fig.~\ref{sgra}. It also includes the expected
$\gamma$-emission from the pion-decay. What makes the $pp$-process so
interesting is that it drops drastically when the temperature
falls below $10^{12}$ K. This is the case at the inner edge of hot
accretion flows when one reduces the spin of the black hole (e.g.,
\cite{Manmoto2000}). As a consequence the jet quickly switches to
a ``radio-quiet'' state when the spin drops below maximal (dashed line
in Figure~\ref{sgra}). Hence, $pp$-collisions are an interesting
mechanism for particle injection into jets, naturally providing a
fundamental switch between radio-loud and radio-quiet jets, and
establishing a link between black hole spin and radio-loudness.

In this picture radio-loud jets would be a mixture of a pair plasma
and a normal plasma and require hot accretion flows around maximally
spinning black holes. The difference in accretion disk structure in
radio-loud/radio-quiet AGN could be reflected in their X-ray spectra
as reported by \citeN{EracleousSambrunaMushotzky2000}.

\section{Summary} 
Let us now summarize the findings of our exploration into the
non-blazar space. First we need to define what we actually mean by BL
Lac or blazar-like when applying these terms to weaker sources. The
main feature of BL Lacs is their dominant non-thermal broadband
spectrum. Domination here needs to be taken relative to other emission
components from the {\it black hole system}, such as the thermal
emission from the accretion flow, and not, e.g., relative to the host
galaxy. This means that classifications based on ground-based
spectroscopy (break contrast, equivalent width) will loose their
usefulness for low AGN luminosity levels
\cite{MarchaBrowneImpey1996}. 

Relativistic boosting is another important feature that characterizes
a luminous blazar. It mainly serves to enhance the non-thermal over
the more isotropic thermal emission. If, however, the thermal emission
is suppressed by other means, boosting may not be so important
anymore. The physics inside the jet, particle acceleration and
non-thermal emission mechanisms, will remain the same with or without
boosting. In fact from viewing BL Lacs or blazars at different angles
and studying them at different power levels and in different
environments we may have yet a lot to learn about them.

To account for these two different definitions, that may not
necessarily always go together, we have tentatively started to use the
word low-luminosity BL Lac in cases where we mean ``non-thermal
dominance'' and the word low-luminosity or radio-weak ``blazar'' when
talking about boosting.

So, where are the low-luminosity BL Lacs and blazars? In radio-quiet
quasars we have now some good evidence for relativistic jets -- the
variability and the superluminal motion. This clearly speaks for the
presence of radio-weak blazars in some of them, such as III~Zw~2. They
will, however, hardly ever appear as BL Lacs, since even if boosted
the non-thermal emission cannot overwhelm the accretion disk emission
that is always there.

On the contrary, we find a number of BL Lac-like sources in the cores
of low-luminosity AGN, such as LINERS, with core dominance and strong
variability. In many cases thermal X-rays are very weak, e.g. in some
ellipticals (Mushotzky, priv. comm.) and it could well be that the
non-thermal emission is the most important. So, they can look like BL Lacs
without necessarily being boosted. This is probably what is seen in
Sgr A* in the Galactic Center and hence we might call it the least
luminous BL Lac.

Finally, we want to mention that another interesting direction to look
in, for future studies, are X-ray binaries. Especially in the
Low/Hard-state, X-ray binaries seem to have very little thermal
emission and show a flat spectrum emission that extends from radio to
optical and then continues in an X-ray power-law.  In trying to fit
the broadband spectrum of the newly discovered X-ray binary XTE
J1118+480 we were able to account for the entire spectrum from radio
to X-rays by emission from a mildly relativistic jet
\cite{MarkoffFalckeFender2000} alone. The question to be asked
therefore is whether this actually is the BL Lac analogue for stellar
mass black holes.



\end{document}